\documentclass{aastex631}

\received{October 7, 2020}
\accepted{July 23, 2021}

\shorttitle{Tying the geometrical traits of MYSOs to a potential evolutionary sequence}
\shortauthors{Frost et al.}
\graphicspath{{./}{figures/}}
\begin{document}

\title{Tying the geometrical traits of massive young stellar objects and their discs to a potential evolutionary sequence using infrared observations}

\correspondingauthor{A. J. Frost}
\email{afrost274@gmail.com}

\author[0000-0001-9131-9970]{A. J. Frost}
\affiliation{Institute of Astronomy, KU Leuven, Celestijnenlaan 200D, Leuven, Belgium}
\affiliation{School of Physics and Astronomy, University of Leeds, Leeds LS2 9JT, UK}

\author{R. D. Oudmaijer}
\affiliation{School of Physics and Astronomy, University of Leeds, Leeds LS2 9JT, UK}

\author{S. L. Lumsden}
\affiliation{School of Physics and Astronomy, University of Leeds, Leeds LS2 9JT, UK}

\author{W. J. de Wit}
\affiliation{European Southern Observatory, Alonso de Cordova 3107, Vitacura, Santiago, Chile}



\begin{abstract}

Young massive stars influence their surroundings from local to galactic scales, but the observational challenges associated with their distance and embedded nature has, until the recent decade, made high-resolution studies of these objects difficult. In particular, comparative analyses of MYSO discs are currently lacking and our understanding of their evolution is limited. Here, we combine the results of two studies with the aim to attribute geometrical features to an evolutionary sequence for a sample of seven MYSOs. The time evolution is based on a near-IR spectral features, while the geometry is determined from a multi size-scale study of MYSOs. We find that MYSO discs with determined geometrical substructure turn out to be also spectroscopically more evolved. This implies that 
%In particular, detailed comparative analyses of the discs of massive young stellar objects (MYSOs) are lacking, and our understanding of MYSO disc evolution is limited. Here, we combine the results of two studies in an attempt to attribute the geometrical characteristics of MYSOs to an evolutionary sequence.  %combines the results of a high-resolution, multi-scale analysis survey of seven MYSOs with previous spectral work to tie the geometrical characteristics of the MYSOs to an evolutionary sequence. 
%By combining a classification scheme derived from variations observed in the  near-infrared spectra of MYSOs and the independent results of a high-resolution, multi-scale survey, %and the geometrical traits were determined through fitting 2.5-3D RT models to N-band interferometric data, Q-band imaging data and spectral energy distributions simultaneously. This is the largest sample of MYSOs for which such a multi-scale detailed analysis has been conducted. 
%we attribute the geometrical characteristics of  MYSOs to their age and find that MYSO discs with substructure and inner clearing are more evolved than those without. This implies that 
disc evolution and dispersal are occurring within MYSOs similar to low-mass YSO disc evolution, despite their faster formation timescales.

\end{abstract}



\section{Introduction} \label{sec:intro}

Massive stars ($\geq$8M$_{\odot}$) are some of the most influential objects within the Universe, affecting their own local environments and the galaxy as a whole. Through their winds and outflows, they mould molecular clouds \citep{krum14} affect further star formation and shape galactic superwinds \citep{leith}. When they end their lives through supernovae events, they synthesise the heaviest elements in the Universe and distribute these throughout the interstellar medium (ISM), enriching the material of future stellar systems like our own. Additionally, when they end their lives they are the only stars to create black holes and as such are the progenitors of gravitational wave sources \citep{gw}. Despite their importance, the formation process of massive stars is not well understood as they are distant, deeply embedded and rare \citep{motte18}. One phase of evolution which is of particular interest to study is the so-called massive young stellar object (MYSO) - an infrared-bright object whose spectral energy distributions (SEDs) peaks in the far infrared  and has a total luminosity around 10$^4$L$_{\odot}$. The nature of MYSOs, in particular at scales of 100s of au which will be involved with important processes such as accretion, has remained elusive. %Previous studies have shown MYSOs to harbour reservoirs of natal material known as envelopes \citep{witcomics} and outflows are detected ubiquitously around these sources . 
Interferometric observations in general provide higher resolution views of massive forming stars than those obtained through other observing techniques, reaching scales of $\sim$0.01". This allows interferometry to probe the innermost regions of MYSOs involved in the accretion process. In the infrared, such interferometric studies have revealed the presence of discs around MYSOs (e.g. \citet{kraus}, 8au spatial resolution) and the presence of close binary systems (e.g. \citealt{koumpia} at $\sim$30au). In the sub-millimetre, ALMA has provided new insights to the environments of MYSOs at $\sim$200au scales (e.g. \citealt{maudafgl1}). Using continuum measurements and molecular studies, structures such as warps (\citealt{sannawarp}, \citealt{monwarp}), spirals \citep{kath20} and rings \citep{maudafgl2} have been detected in MYSO discs. The presence of such discs is of great interest to probe the way MYSOs accrete mass, as for low-mass stars, disc-accretion is the main avenue of mass transfer from the wider environment to the star. For low-mass stars, large interferometric surveys of discs have been performed, allowing detailed investigations of their composition and the processes occurring within them (e.g. DSHARP, \citealt{dsharp}). These types of surveys, combined with previous works which defined evolutionary classes of low-mass discs based on their spectral SED emission \citep{bars, shu77}, have allowed different disc features to be attributed to different ages of protostellar disc. Younger discs show evidence of accretion features in their SEDs and their minimum dust radius is often dictated by dust sublimation alone \citep{bdw}. On the other hand, inner clearing is observed in so-called `transition discs' \citep{cieza}, one elder form of disc. Other structures such as spirals and rings in discs are typically observed in elder sources, save for a few examples (e.g. \citealt{sheehan}). Studying such features is critical in determining the physical processes occurring with the disc, such as those related to planet formation, dust amalgamation or the formation of companions. Such a classification of the evolution of MYSO discs has, thus far, not been feasible. 

In \citet{frost2}, we used a approach which used multiple datasets which trace 10mas scales, 100mas scales and SEDs to characterise the geometry of a sample of eight MYSOs but with no information on their age. Independently, \citet{coop} and \citet{coop13} studied a sample of nearly 130 MYSOs using their near-infrared colour and near-infrared spectra. In doing so, they noticed different spectral lines in the redder, more embedded and therefore younger sources to the elder bluer sources and from this developed a classification scheme associated with evolutionary phase. In this paper, we combine the results and methods of these two studies to attribute the geometrical characteristics of MYSOs (and specifically their discs) to a potential evolutionary sequence for the first time. In the next subsections, we introduce the methods and results of \citet{frost2} and \citet{coop} for ease of interpretation of the results of this paper. In Section 2, we present the evolutionary classification of the sources from \citet{frost2} and use their geometrical features to compare to previous star formation models and estimate numerical ages for the sources. In Section 3, we discuss our results before concluding. %\textbf{We employed their classification scheme to the sample of MYSOs studied in \citet{frost2}. Spectra to perform the classification were obtained from the literature for all sources in \citet{frost2} save one, therefore leaving a sample of seven sources.} In combining these two works for these objects , we find a natural consistency between the sub-arcsecond MYSO geometry and their emitted near-IR spectra. We use this to tie the geometrical characteristics of the MYSOs to a potential evolutionary sequence, with the most marked variation across the different stages observed in the object's discs. \textbf{The sources we classify are listed in Table \ref{mysosampletypes}.:}

\subsection{Work thus far I - Characterising the geometry of MYSOs at multiple scales}

The most detailed interferometric studies of MYSOs tend to focus on single MYSOs. Such individual studies are not homogeneous, as they use different methods and types of analysis. \cite{boley13} presented a survey of 20 MYSOs with the now-decommissioned VLTI MIDI instrument fitting geometric models to 10.6$\mu$m data, providing a first glimpse of mid-IR structure at 100au scales around MYSOs and attributing this to disc and outflow emission. Some studies have investigated MYSOs using their spectral energy distributions (SEDs) which provide information on the whole source, but with no direct spatial information. In \citet{mengyao}, the model grid of \citet{zt} was used to fit SEDs of a number of MYSOs. However the interpretation of SED data can be difficult in particular given the strong dependencies on parameters such as viewing angle (e.g. \citealt{whit03}, \citealt{rob17}) and the embedded nature of MYSOs means that disc emission is not easily distinguished from stellar/envelope emission as can be done for low-mass stars. \citet{witcomics} and \citet{wheel} fit Q-band image profiles and SEDs using 2D radiative transfer (RT) models, therefore linking the information of an SED to a high $\sim$0.1" resolution dataset but lacking the sensitivity to small spatial scales that interferometry could provide. \citet{frost} and \citet{frost2} combined all these observables, using a detailed methodology which combines the N-band ($\sim$7-13$\mu$m) interferometric data from \citet{boley13} (required to probe the smallest 0.01" spatial scales) with near-diffraction-limited imaging in the Q-band (19.5$\mu$m - 24.5$\mu$m) from \citet{frost2} and \citet{witcomics} and SEDs. The SEDs were compiled using the online database of the Red MSX Source (RMS) survey and the literature, including fluxes from instruments such as 2MASS, WISE, GLIMPSE, Herschel/PACS, ATLASGAL, SCUBA and more (see \citealt{frost2} for a complete list) covering near-infrared to millimetre wavelengths. These observables were fit with synthetic observations from RT models using the 3D dust RT code HOCHUNK-3D \citep{whitney}. This code simulates the protostellar environment through the combination of different geometrical components; a central source defined by its mass, temperature and radius, two disc components (a `large grains' disc and a `small grains' disc each with their own determinable scale-heights, minimum radii and maximum radii) that contribute to one overall disc mass, a surrounding envelope component and bipolar outflow cavities. The models from the RT code were fit to the three observables simultaneously. Synthetic images at 1$\mu$m intervals were generated in the N-band and post-processed to produce visibilities at each wavelength to be compared to those observed \citep{frost}. Similarly Q-band synthetic images were generated by the code, convolved to match the observations and radial profiles extracted from them to allow comparison to diffraction-limited $\sim$20$\mu$m images. In doing so the characteristics of a sample of MYSOs were obtained at multiple scales and the combination of high-resolution datasets allowed the distinction between disc emission and other components.

In \citet{frost2} we applied this methodology on a sample of eight MYSOs, the largest sample to be analysed in such a way. The sources were selected randomly based on the availability of both archival interferometric N-band data (MIDI, \citealt{boley13}) and the additional presence of Q-band imaging data \citep{frost2, witcomics}. When fitting the Q- and N-band datasets simultaneously, the cavity wall geometry was constrained at two scales. With this in place, it was found that one could not satisfy the N-band and Q-band data simultaneously without including discs in the models (see Figure 8, \citealt{frost}). The simulated N-band emission is affected most by the presence of a disc, with the N-band visibilities higher in a disc and envelope model than an envelope-only model due to the presence of a compact object. The geometry of the disc is also significant, with the simulated visibilities particularly sensitive to disc inner radius and scale-height in particular, with disc outer radius and total mass being less well constrained. For one source (NGC 2264 IRS1) which is close to face-on, the mass was better constrained as the disc data was best fit with a mass-dependent spiral structure. The SED provides additional information on the wider cooler environment out to millimetre wavelengths. With these three observables combined, a multi-scale picture of the MYSOs is obtained and a wide parameter space encompassing the discs, cavities and envelopes of the sources can be investigated.%By fitting 2.5-3D radiative transfer models (HO-CHUNK3D, \citealt{whitney}) with free parameters covering disc, central star, cavity and envelope characteristics to the combined datasets, 

In \citet{frost2} we concluded that, in agreement with previous works, the geometry of these MYSOs consisted of dusty envelopes of natal material which are rotating and infalling (following the solution of \citealt{ulrich}). Within these envelopes bipolar outflow cavities are present, carved out of the natal material by outflows, which are ubiquitously observed around MYSOs. Importantly, discs are present at scales of $\sim$100au, which could therefore be involved in the accretion process. All sources in the sample are all relatively similar at large scales, with envelope radii of the order 10$^5$au from the fitted models. Similarly the envelope infall rates of the sample were all $\sim$10$^{-4}$M$_{\odot}$yr$^{-1}$ (except for one source (S255 IRS3) which is known to be episodically accreting \citep{car18} and for which the data used were from an accretion lull for the source). A range of opening angles (measured from the polar axis) and cavity densities were seen across the sample with minima and maxima of 12$^{\circ}$ and 40$^{\circ}$ and $\sim$10$^{-19}$gcm$^{-3}$ and $\sim$10$^{-21}$gcm$^{-3}$ respectively. The best-fit models of the discs in each source vary considerably, some are flat (scale-height $\sim$0.1) while others are flared (scale-height $\geq$0.4). Finally, it was found that adding substructure the discs in the models improved the fits to the observational data. Most commonly this substructure takes the form of cleared inner regions, where the minimum dust radius is greater than the sublimation radius of the central source. The best-fit model for one source (NGC 2264 IRS1) incorporates a spiral/gap structure within its flared disc.

%In this letter, we bring the results of this multi-scale sample together with previous spectral work, to demonstrate how these geometric characteristics align with a potential evolutionary sequence and present a consistent comparative analysis of the MYSOs.

  \begin{table*}
\caption{In this table we provide background information on the MYSOs we classify in this study to put them into context in the wider MYSO population. %We also include a few key parameters from our previous study of the geometry of the MYSOs \citep{frost2}.
 \newline References: (1) \citet{rmslum}; (2) \citet{rmsmott10}; (3) \citep{mottlum}; (4) \citet{immer}; (5) \citet{grell}; (6) \citet{kamezaki}; (7) \citep{s255l}; (8) \citet{s255d};
%(9) \citet{hen92}; (10) \citet{mondist}; 
(9) \citet{kraus17}; (10) \citet{boley13}; (11) \citet{linz09}; (12) \citet{damgaia}; (13) \citet{urq12}; (14) \citet{urq14}; (15) \citet{frost2}.}         % title of Table
\label{mysoinfo}      % is used to refer this table in the text
\centering
%\begin{adjustbox}{width=1\textwidth} 
\hspace*{-2cm}\begin{tabular}{c c c c c}          % centered columns (4 columns)
\hline%\hline                        % inserts double horizontal lines
Name & RA (J2000) & DEC (J2000)& log($\frac{L}{L_{\odot}}$) & Distance \\ % & Substructure?\textsuperscript{15} & Cavity opening angle\textsuperscript{15} & Cavity density\textsuperscript{15}\\ 
 & (h:m:s) & (d:m:s) & &(kpc) \\ % & ($^{\circ}$) & (gcm$^{-3}$) \\
%& & & & & COMICS (C) data? \\% table heading  % table heading
\hline                                   % inserts single horizontal line
G305.20+0.21 & 13:11:10.45 & -62:34:38.6 & 4.7\textsuperscript{1} & 4.0\textsuperscript{2} \\ %& Cleared inner disc regions \\
W33A & 18:14:39.0 & -17:52:03 & 4.5\textsuperscript{3} & 2.4\textsuperscript{4} \\ %& n/a \\
NGC 2264 IRS1 & 06:41:10.15 & +09:29:33.6 & 3.6\textsuperscript{5} & 0.7\textsuperscript{6} \\ %& Mass-heavy spiral structure \\
S255 IRS3 & 06:12:54.02	& +17:59:23.60 & 4.7\textsuperscript{7} & 1.8\textsuperscript{8} \\ % & n/a \\
%Mon R2 IRS2 & 06:07:45.8 & -0.6:22:53.2 & 3.8\textsuperscript{9} & 0.8\textsuperscript{10} & III \\
IRAS 17216-3801 & 17:25:06.51 & -38:04:00.4 & 4.8\textsuperscript{9} & 3.1\textsuperscript{10} \\ %& Cleared inner disc regions \\
%& & & & & low densities & \\
M8EIR & 18:04:53.18 & -24:26:41.4 & 3.8\textsuperscript{11} & 1.3\textsuperscript{12} \\ %& Cleared inner disc regions \\
%& & & & & low densities & \\
AFGL 2136 & 18:22:26.38 & -13:30:12.0 & 5\textsuperscript{1} & 2.2\textsuperscript{13,14} \\ % & Cleared inner disc regions \\
\hline                                             %inserts single line
\end{tabular}
%\end{adjustbox}
\end{table*}

\subsection{Work thus far II - Spectral variance in MYSOs}

\citet{coop} and \citet{coop13} examined the NIR spectra of over one hundred MYSOs from the RMS survey. Various emission signatures were observed throughout the objects, allowing them to be categorised into three main different types. The emission lines which allowed this were H$_2$, Br$\gamma$, Br10 and fluorescent Fe II (hereafter fl-FeII). %According to Cooper's classification, Type I MYSOs display strong H$_2$ emission lines but no others. Type II have H$_2$ and Br$\gamma$ present, weak or absent Br10 emission and absent fl-FeII. Type III show weak or absent H$_2$, strong Br$\gamma$ emission and a presence of Br10 and fl-FeII emission.

Based on the presence of these lines, the MYSOs were grouped into different evolutionary "Types". Each of these types is classed as a different evolutionary phase. Type I MYSOs display strong H$_2$ (2.1218$\mu$m) emission lines, attributed to shocks from the collision of high-velocity ejections with ambient material \citep{mars}. Type I sources do not display any other lines and are categorised as the youngest sources. Type II sources display H$_2$ and Br$\gamma$ emission, with the latter arising due to the increasing central temperature and luminosity of the central protostar \citep{coop}. Br10 emission is weak or absent, and there is also a lack of fl-FeII emission as the star is not yet evolved enough to produce the ionising photons that cause its emission \citep{coop}. Type III display weak or absent H$_2$ emission, strong Br$\gamma$ emission and a presence of Br10 and fl-FeII emission. As the MYSO evolves through the Type II and Type III stages the amount of H$_2$ emission decreases, perhaps because the outflow itself is weakening, the amount of circumstellar material is depleting due to dispersion and/or because the molecules of H$_2$ are being dissociated as the protostar gets hotter. This increase in temperature also induces (Type II) and increases (Type III) the Br$\gamma$ emission. By the time an MYSO is at Type III, either the outflow is too weak or there is not enough material to form shocks of H$_2$. Past this point ionisation would be prevalent enough to create a H\textsc{ii} region. %Type I would correspond to the youngest, with the strong H$_2$ emission attributed to shock activity where outflow material and envelope material collide. As the MYSO evolves through the Type II and Type III stages the amount of H$_2$ emission decreases, perhaps because the outflow itself is weakening, the amount of circumstellar material is depleting due to dispersion and/or because the molecules of H$_2$ are being dissociated as the protostar gets hotter. This increase in temperature also induces (Type II) and increases (Type III) the Br$\gamma$ emission. By the time an MYSO is at Type III, either the outflow is too weak or there is not enough material to form shocks of H$_2$. At this stage the star would be hot enough to drive a UV wind, generating HI and fl-FeII emission at 1.6878$\mu$m. Past this point ionisation would be prevalent enough to create a HII region. %The object would no longer solely be an MYSO. 

Mid-infrared colour-colour cuts can be used as crude estimators of age and have allowed MYSOs to be distinguished from other evolutionary stages of massive star formation \citep{rmslum}. Within the sample of \citet{coop}, the Type III sources were found to be bluer than the Type I sources, indicating that Type I is the youngest and that the sources may evolve through the Type II stage to the Type III stage. Additionally, He I emission, which is a common signature of main-sequence wind activity, are not found in the Type I sources \citep{coop}. Some presence in Type II sources is observed and He I emission is most prevalent in the spectra of the Type III sources, thereby implying that the Type III are closest to the main-sequence and the most evolved.

For the sources of \citet{frost2} we compiled a catalogue of all spectra required to categorise them according to the outline above. Spectra for 6 of the 8 sources from \citet{frost2} are available from previous works on the RMS survey MYSOs (\citealt{sausage}, \citealt{coop13} and \citealt{coop}). In addition, a spectra of M8EIR was obtained from \citet{porter} and a 2$\mu$m spectra for IRAS 17216-3801 obtained from \citep{kraus17} were sufficient to provide a reasonable characterisation according to \citet{coop}. Therefore, in total, the sample analysed in this paper consists of 7 sources. The sources are listed in Table \ref{mysoinfo}.

\section{New Results} \label{sec:Results}

We present the classifications of the sources in Table \ref{mysotypes}. %Three sources are classified as Type IIIs, two as Type II and one as Type I. AFGL 2136 could be classified as either a Type II or Type III source. ccording to the classification as it lacks H$_2$ emission but only shows weak Br-$\gamma$ emission. \citet{ind20} and \citet{barr2020} studied AFGL 2136 spectroscopically with VLT+CRIRES, SOFIA+EXES, and Gemini North+TEXES spectra. They find that the weak Br-$\gamma$ emission has to be limited to a very small emitting region in the inner disc based on the presence of spatially unresolved H30$\alpha$ emission and  also note that the close to edge-on inclination of the source could lead to the weak appearance of the Br-$\gamma$ line. The presence of the H30$\alpha$ recombination line also implies the presence of ionisation and therefore UV photons, which again implies the central protostar is more evolved and that AFGL 2136 is closer to a Type III than a Type II source.
Most of the sources were straightforward to classify. We classify S255 IRS3 as a Type I source, in agreement with the previous classification by \citet{coop}. This implies that it is the least evolved source in the sample. G305.20+0.21 and W33A are both classified as Type II sources. %Ultimately, we classify the source as a Type II/III and this is discussed further in Section 5. 
NGC 2264 IRS1 and M8EIR are classified as Type III sources.  While the available spectra for IRAS 17216-3801 did not cover the Br series, an absence of H$_2$ emission and the strength of the Br-$\gamma$ line suggests that this is a Type III source, and we discuss the source as such. %Figure \ref{coopevo}  displays how the geometry determined through the multi-scale analysis varies with their derived Types.
One source, AFGL 2136, was more complex to classify. AFGL 2136 could be interpreted as either a Type II or Type III source which presents discussion. Its spectrum displays a lack of H$_2$, implying it is an older Type III object but Br-$\gamma$ is only weakly observed, implying instead it is be a younger Type II object. %Some CO lines are present, but most of these are absorption lines. \citet{hoff} find CO absorption in a number of MYSOs in the M17 region and attribute this to heavy accretion ($>$10$^{-5}$Myr$^{-1}$, consistent with the final model rate found of $\sim$10$^{-4}$Myr$^{-1}$). 
Emission lines like Br-$\gamma$ and H$_2$ could appear weak if the source has a high dust column density, which would increase the amount of continuum emission making the lines look fainter and could be the source of this confusion. The continuum emission of AFGL 2136 is high but its column density was not directly measured in our work. AFGL 2136 could be classified as either a Type II or Type III according to the classification as it lacks H$_2$ emission but only shows weak Br-$\gamma$ emission. \citet{ind20} and \citet{barr2020} studied AFGL 2136 spectroscopically with VLT+CRIRES, SOFIA+EXES, and Gemini North+TEXES spectra. They find that the weak Br-$\gamma$ emission has to be limited to a very small emitting region in the inner disc based on the presence of spatially unresolved H30$\alpha$ emission and  also note that the close to edge-on inclination of the source could lead to the weak appearance of the Br-$\gamma$ line. The presence of the H30$\alpha$ recombination line also implies the presence of ionisation and therefore UV photons, which again implies the central protostar is more evolved and that AFGL 2136 is closer to a Type III than a Type II source.

\subsection{Comparison with previous 3D radiation-hydrodynamical simulations}

Following the determination of a potential evolutionary phase for each of the MYSOs, we additionally considered the work of \citet{off}. These authors investigate how protostellar outflows change for YSOs with 3D radiation-hydrodynamic simulation.%, building on prior hydrodynamical studies of turbulent stellar clusters . 
In the simulations, a turbulent steady state was achieved and the gravitational energy is comparable to the turbulent energy in the box at $t$=0 \citep{off2}. Energy is injected into the simulation and after the initial driving phase the turbulence decays on a dynamical timescale, initiating a global collapse. Protostars are added to this initial space which then generate outflows characterised by their ejection efficiency, velocity and momentum distribution. Using these models, \citet{off} find that the cavity opening angles of the YSOs increase with age. Similar results have been found in works such as \citet{beutshep} and \citet{vaidya}.

Given that the cavity opening angle was a parameter which could affect both the N- and Q-band simulated high-resolution datasets throughout our previous work in \citet{frost2}, it was of interest to see what ages \citet{off} would associate with the final geometries derived in \citet{frost2} and to see how these compared to the evolutionary Types derived for the sources in this paper. For this reason, we reiterate some relevant characteristics from \citet{frost2} in Table \ref{mysotypes}. It must be noted that in \citet{frost2} and here what we refer to as the `cavity opening angle' is the cavity `half angle' in \citet{off}. Henceforth we continue to use 'cavity opening angle' in consistency with our previous work. The cavity opening angle of S255 IRS3 is one of the smaller of the sample at 20$^{\circ}$. This corresponds to a protostellar age of $\sim$10$^{3}$yrs according to the modelling of \citet{off}. W33A also has a 20 degree cavity putting it at the younger end of the evolutionary track of \citet{off}. The cavity opening angle for G305 is the smallest of the sample, but this value may be affected by a secondary dusty object detected in the cavity by \citet{mengyao} (see Paper I for further discussion). As a result we do not calculate an age for this source. AFGL 2136 can, as previously discussed, be considered either a Type II or Type III source and its cavity opening angle lies between those determined for W33A (Type II) and the Type III sources. The cavity opening angles for the Type III sources are also the larger of the sample, and according to \citet{off} they range in age between $\sim$10$^4$-10$^5$ years old. Therefore, this comparison provides independent support of the classification using the Types based on the near-infrared spectra, with the elder Types having the largest ages, though given our low-number statistics further investigation is required.

  \begin{table*}[h]
\caption{In this table we present the types of the sources we derive for the MYSOs of our sample and the ages we derive by comparing our cavity sizes to the work of \citet{off}. $\theta_{cav}$ is the cavity opening angle. The units of the line fluxes are Wm$^{-2}\mu$m$^{-1}$. For clarity, we also include the emission line information from \citet{coop}, \citet{sausage}, \citet{porter}and \citet{coop13} used in the classification process and some relevant geometric information from our previous work, \citet{frost2}, which we discuss in Section 4. The cavity opening angle of G305 is starred as this sources has a secondary dusty source within its outflow cavity, which likely affected the fitting process. As a result, we do not calculate an age for this source. A full discussion can be found in \citet{frost}. The spectral information for M8EIR comes from \citet{porter}, where the line strength is presented relative to the flux of the Br$\gamma$, so these ratios are reiterated for that source. No line data are included for IRAS 17216-3801 as the emission lines were present in the differential phases and visibilities of the spectrointerferometric data of \citet{kraus17}. The non-dectection of H$_2$ for AFGL 2136 is discussed in Section 2.}
\label{mysotypes}      % is used to refer this table in the text
\centering         
\begin{adjustbox}{width=1\textwidth}
\hspace*{-2cm}\begin{tabular}{c c c c c c c c c c}          % centered columns (4 columns)
\hline%\hline                        % inserts double horizontal lines
Name & R$_{min}^{dust}$ & $\theta_{cav}$ & Cavity density & Type & H$_2$ flux & Br$\gamma$ flux & Br 10 flux & fl-FeII flux & Age \\
& (au) & ($^{\circ}$) & (gcm$^{-3}$) & & & & & & (log yr) \\ 
%& & & & & COMICS (C) data? \\% table heading  % table heading
\hline  
S255 IRS3 & 12 (=R$_{sub}$) & 20 & 6$\times$10$^{-19}$ & I & (1.20$\pm$0.06)$\times$10$^{-17}$ & $<$1.2$\times$10$^{-17}$ & $<$2.5$\times$10$^{-18}$ & $<$2.1$\times$10$^{-18}$ & 3.1\\% inserts single horizontal line
W33A & 18 (=R$_{sub}$) & 20 & 1$\times$10$^{-19}$ & II & (8.7$\pm$0.1)$\times$10$-18$ & (2.1$\pm$0.2)$\times$10$^{-17}$ & - & - & 3.1 \\
G305.20+0.21 & 60 ($\sim$3.5R$_{sub}$) & 12* & 1$\times$10$^{-19}$ & II & (4.2$\pm$0.003)$\times$10$^{-17}$ & (1.1$\pm$0.002)$\times$10$^{-17}$ & - & - & - \\
AFGL 2136 & 125 ($\sim$4R$_{sub}$) & 22.5 & 3$\times$10$^{-19}$ & II/III & - & (2.9$\pm$0.1)$\times$10$^{-17}$ & - & - & 3.5\\
NGC 2264 IRS1 & 4 (=R$_{sub}$) & 25 & 8$\times$10$^{-21}$ & III & $<$7.0$\times$10$^{-16}$ & (9.9$\pm$0.3)$\times$10$^{-16}$ & (4.0$\pm$0.3)$\times$10$^{-16}$ & (1.8$\pm$0.3)$\times$10$^{-16}$ & 3.8 \\
%Mon R2 IRS2 & 06:07:45.8 & -0.6:22:53.2 & 3.8\textsuperscript{9} & 0.8\textsuperscript{10} & III \\
IRAS 17216-3801 & 100 ($\sim$3R$_{sub}$) & 40 & 9$\times$10$^{-21}$ & III & - & - & - & - & 5.0 \\
%& & & & & low densities & \\
M8EIR & 30 ($\sim$1.5R$_{sub}$) & 25 & 8$\times$10$^{-21}$ & III & - & 1 & 0.37 & 0.06 & 3.8 \\
%& & & & & low densities & \\

\hline                                             %inserts single line
\end{tabular}
\end{adjustbox}
\end{table*}

%   \begin{figure}
%   \centering
%   \includegraphics[width=180mm]{cooperevo.pdf}
%   \caption{Diagram showing how the geometry of the MYSOs varies according to the Types %of \citet{coop}. We include the example spectra of each Type from their work for %completeness.}
%   \label{coopevo}
%   \end{figure}

%A graphical presentation of how the parameters of the final models for the sample vary with their classified type is shown in the Appendix (Figure \ref{typesimg}).

   \begin{figure}
   \centering
   \includegraphics[width=180mm]{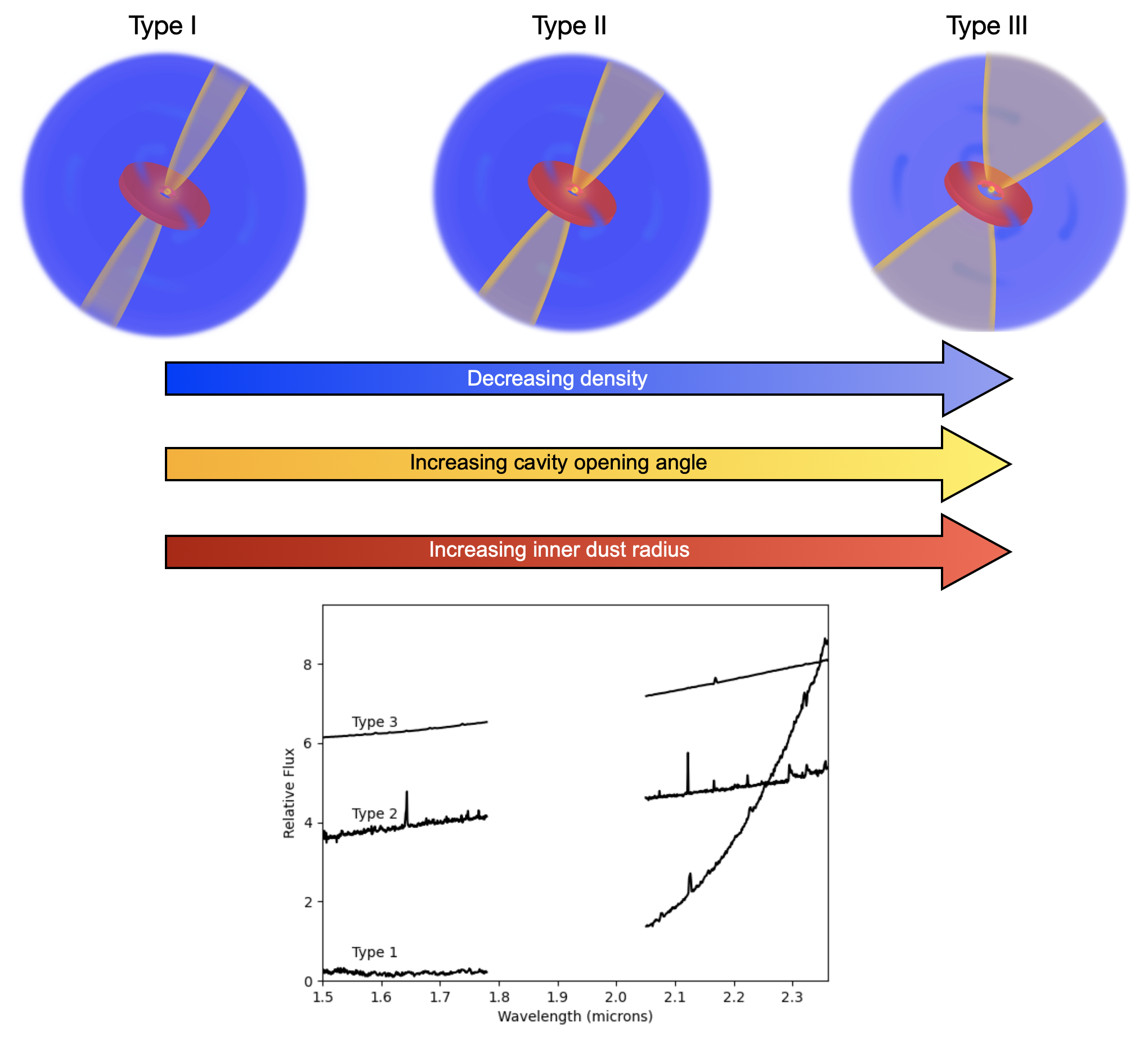}
   \caption{\textit{Top}: Diagram showing how the geometrical traits derived for the sample of MYSOs vary according to the evolutionary Types. As the Type increases, the cavity densities decrease, the cavity opening angles widen and the disc regions clear (the inner radius increases), as highlighted by the arrows. \textit{Bottom}: Example spectra of each Type. The Type I MYSO is S255 IRS3, the Type II MYSO is G305.20+0.21 and the Type III MYSO is NGC 2264 IRS1. %\textit{Bottom right}: Histogram showing whether substructure is present with the MYSOs for each evolutionary type. \textbf{Within this work we define substructure as a) a deviation from axisymmetry within the disc or 2) the expansion of the inner dust radius beyond the dust sublimation radius).}%Diagram showing the typical geometry adopted for the MYSOs in this work. This consists of a central disc, which can have substructures, surrounded by a dusty envelope. In the polar directions, perpendicular to the disc plane, there are cavities within the envelope carved by outflows. The N-band emission originates from the very central hot regions, while Q-band imaging probes larger scales more sensitive to outflow cavity walls. Photometric observations used to build the SED are low resolution and sensitive to the entire source structure. As such they are dominated by envelope emission.
   }
   \label{coopevo}
   \end{figure}

\section{Notable trends \& Discussion} \label{sec:discussion}

The majority of the  Type III sources have lower densities in their cavities ($\sim$10$^{-21}$gcm$^{-3}$, \citealt{frost2}). They show the widest cavity opening angles of the sample with a maximum angle of 30$^{\circ}$. The modelled discs providing the best fit to the Type III sources are all required to have some form of substructure, generally an inner clearing. NGC 2264 IRS1 is the only source whose model shows a substructure other than an inner hole in the form of a gap-like structure \citep{frost2}. 
For low-intermediate mass stars such a feature is thought to form slowly through dust growth (e.g. \citealt{zhang15}) or binary interactions (e.g. \citealt{price18}) and are expected for more evolved sources. Typical timescales of disc evolution in low-mass star formation are on the order of mega-years. The later timescales of disc evolution are characterised by the presence of  substructures like rings and gaps. One suspected origin of such structures is dust trapping, where grains collect in pressure minima or maxima throughout the disc. Such trapping gathers grains in close proximity where they can amalgamate and grow \citep{wcrev}. However, recent observational studies are providing evidence for dust trapping occurring much earlier. For example, \citet{sc20} find evidence of four annular gap-like structures in the source IRS 63 which is less than half a million years old. These authors state posit that such structures are an area of likely grain growth. The case of IRS 63 is particularly relevant to comparison with the MYSO case, as it is a so-called Class I source \citep{shu87} which, like an MYSO, is expected to have not only a disc but also surrounding envelope material. Additionally, the age of IRS 63 is within predicted massive star formation timescales \citep{tanmsf}. Thus, if grain growth is potentially occurring in a low-mass source, whose dust reservoirs and gravitational potentials are significantly smaller than those that would be expected for a massive protostellar system, it is not unreasonable to postulate that similar annular structures could form within the discs of MYSOs.

The other cause of gaps and rings in low-mass discs is the presence of companions. Recent modelling of massive protostellar discs specifically by \citet{oliva} shows that such spiral substructures can be expected to form for MYSOs too. The disc fragments as the MYSO evolves and eventually forms a binary system. The most complex spiral structure found in their work appears between 4000-15000yrs before a quiescent phase where the spiral structure is more muted. While the source which was best fit with a spiral structure (NGC 2264 IRS1) has not been confirmed to be a binary, the expected high binary fraction for massive stars at both the MYSO (e.g. \citealt{sausagebinary}, \citealt{koumpia}, \citealt{meyerbinary}) and evolved stages (e.g. \citet{sana}), combined with the fact that we see structures known to be induced by companions through modelling (e.g. \citealt{oliva}, \citealt{price18}) in what we classify as an evolved disc is encouraging. 

Another notable trend amongst the Type III discs specifically is that they are all highly flared. Varying disc flaring was required to improve the fit to the near-infrared shoulder of the SEDs and to improve the fit to the MIDI visibilities. A decrease in disc flaring has been presented as a diagnostic for evolution with the low-mass disc community, but the topic remains controversial. For example \citet{meeus}, through the study of infrared excess in the SEDs of Herbig discs, concluded that the earlier discs (dubbed Group I discs) were more likely to be flared and the second group (Group II) were more likely to be flat, attributing the change to dust growth. However, recent high-resolution imaging by \citet{garufi} found that within a sample of 17 Herbig discs (half of all known at the time) that over half the flared discs are older than 10Myr, stating that the idea of evolution from flared to flat discs may need to be revised. The flaring of the disc could then be due to some other characteristic of the protostellar environment. The presence of flared discs in our more evolved MYSO sources is consistent with these findings that flared discs can be found even at later evolutionary stages in low-mass sources.

The Type II sources within this sample display very different geometries. G305 has a large inner hole and a low density environment similar to the more evolved Type III sources. The nature of W33A is less clear. %Investigations into W33A with sub-mm data show the region to be complex with a large-scale spiral-like structure observed with ALMA \citep{maudw33a}, for example. %W33A was previously studied in detail in \citet{wit10} using only MIDI data and SED. It was found that an envelope-cavity model was sufficient to fit the data with cavity opening angles of 10$^{\circ}$. When the source was revisited this geometry was revised - a disc was detected around the source and the cavity opening angle was determined to be 20$^{\circ}$. As discussed in \citet{frostthesis}, the differences between the results are likely due to a more accurate distance of 2.4kpc being determined for W33A since the publication of the findings of \citet{wit10}. 
Modelling work by \citet{izq18} find that the region is very active, supporting the idea that this source may be young amongst the wider population of MYSOs. The fact that the final model of W33A has a disc with no substructure also supports that this may be a younger source, if it is assumed that substructure appears with evolution as it does for low-mass stars. Alternatively, \citet{vdt05} note the presence of multiple millimetre sources in their work at 3000-5000au separations. Given the shorter wavelength range of our high-resolution datasets we do not probe the outer, cooler regions of the disc in detail. These multiple sources may explain the complexity of the region seen with ALMA \citep{maudw33a} but do not assist with the interpretation of our single-source models of the observed data. %On the other hand, previous work by \citet{vdt05} presents a possible detection of a compact HII region, thereby implying ionisation and an older source. 
 %This complex nature explain why W33A shows characteristics of both a old and young source. %This may be expected, as Type II is essentially a go-between from the non-dispersed Type Is to the highly dispersed Type III sources, and while the classes are discrete, the actual evolutionary process of massive protostellar discs should be a continuous sequence. As such we conclude that W33A is a younger source within the Type II classification, while .
%These were inspected to see if certain geometries and characteristics were typical of MYSOs within each of these types. 
%All the discs within this type display substructure.

S255 IRS3 has been shown to be actively accreting \citep{car18} and this corroborates the young classification made from its spectra. The evolution of discs around low-mass stars is marked by inside-out dispersal. This can be initiated by photoevaporation or the presence of a secondary body, which causes the inner regions of protoplanetary discs shift from a continuous structure to a more radially concentrated one \citep{wcrev}. The inner geometry of the disc of S255 IRS3 supports the postulation that this is a young source, as it is not one of the sources which displays inner clearing - the minimum dust radius is the sublimation radius. %\textbf{The final disc still harbours small grains adding further support to the classification, as these would the first grains to be dispersed from the environment.%, but the flat nature of both the small and large grains discs is not suspected of young discs (at least based on observations of low-mass YSOs). 
The envelope infall rate of S255 IRS3 may imply an older source, but the data used to determine the source's geometry are from an accretion lull. Thus, this is likely just a result of the confirmed variable accretion associated with the source. The cavity density ($\sim$10$^{-19}$gcm$^{-3}$, \citealt{frost2}) for the source is one of the higher amongst the sample and again supports the younger age found for the source, as less envelope dispersal would have occurred.% a lack of dispersal of the wider environment and, again, a younger age for the source.  %The fact that S255 has undergone a violent accretion burst may assist with explaining why this source's cavity opening angle is larger than expected. Should this accretion be occurring periodically, a previous accretion-burst event could have disrupted a wider area of the envelope material, leaving a wider cavity which is still being illuminated despite the fact the source is in an accretion lull and could be young. 
The differences observed between the MYSOs with Type is illustrated in Figure \ref{coopevo}. To summarise, cavity opening angle appears to increase with evolutionary stage whilst the overall density of the dust environment decreases and substructure in discs appears over time, being most prevalent in the eldest classified discs. 

   \begin{figure}
   \centering
   \includegraphics[width=80mm]{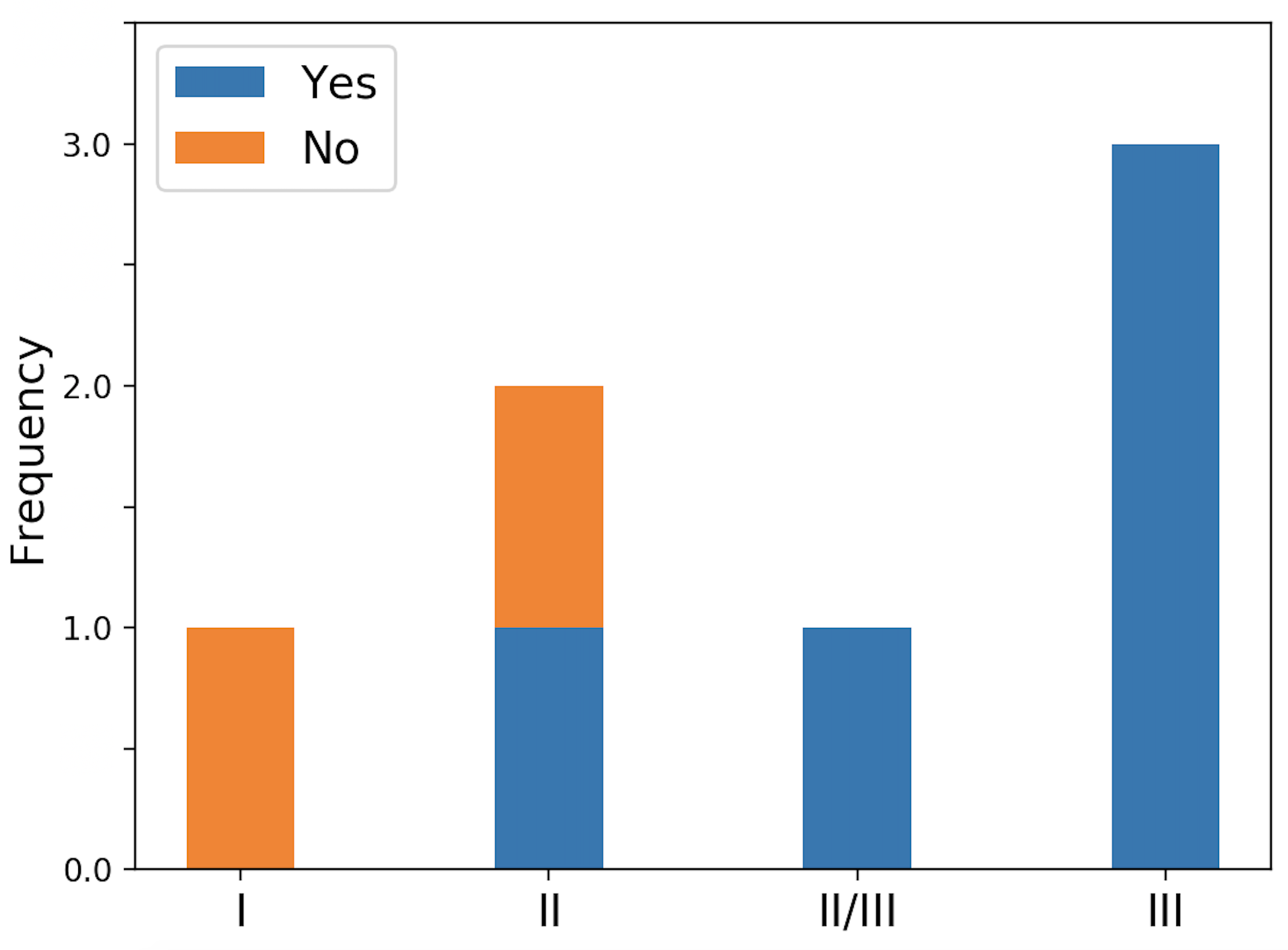}
  \caption{Histogram showing whether substructure is present with the MYSOs for each evolutionary type. Within this work we define substructure as a) a deviation from axisymmetry within the disc or 2) the expansion of the inner dust radius beyond the dust sublimation radius).}
   \label{clearinghist}
   \end{figure}

%   \begin{figure}
%   \centering
%   \includegraphics[width=90mm]{RminsandL.pdf}
%   \caption{Scatter plot showing the difference between $R_{sub}$ (orange) and the minimum radius of the disc (blue).}
%   \label{rminrsubcomp}
%   \end{figure}

\subsection{What is causing the inner clearing in these MYSOs?}

Within our sample, the presence or absence of substructure seems to be associated with evolutionary Type. The histogram in Figure \ref{clearinghist} displays how the presence of substructure varies with type. Two kinds of substructure are inferred from the fitting; inner holes and one case of spiral/gap structure. In order for the inner holes to be present, something must be disrupting the dust beyond the sublimation radius. %Figure \ref{rminrsubcomp} compares the minimum disc radius with that of the sublimation radius of the source. %A strong correlation (0.92, Paper II) exists between disc minimum radius and luminosity. At a first glance, this appears consistent; a more luminous star will be able to sublimate dust at larger distances compared to a less luminous source, increasing the minimum radius of the circumstellar material. However, the minimum radius quoted here is not the dust sublimation radius for the majority of these sources, as illustrated in.
A number of theories exist as to what causes inner clearing in low-mass protostellar discs; photoevaporation \citep{armie}, viscous evolution \citep{esptrans}, the presence of binary/multiple companions \citep{price18} and the presence of planets \citep{dhruv}. For massive stars, modelling works such as \citet{meyer} show that gaps can evolve around massive sources. Photoevaporation and the presence of a binary companion appear to be the best candidates for these MYSOs, given their large luminosities and the high multiplicity fraction expected for massive stars. A calculated Pearson correlation coefficient of 0.93 (a measure of linear correlation calculated as the covariance of two variables divided by the product of their standard deviations) between the luminosity and the model minimum dust radius indicates a strong correlation. Since the minimum dust radius is often larger than sublimation radius, this could suggest that photoevaporation is the mechanism that is disrupting the dust. The classified types of the MYSOs also appear to support that photoevaporation is the cause of this clearing. Photoevaporation occurs due to the presence of UV photons. When an MYSO starts producing UV photons it passes into its next stage of evolution, the ultra-compact H\textsc{ii} region. One of the defining features of the Type III sources is that they display emission lines associated with photoevaporation. As a result it is consistent that more evolved MYSOs are starting to generate a surplus of these disrupting photons and therefore that this is the cause of the clearing. As the Type III discs are also all flared, also lends support to photoevaporation being present and a mechanism of dispersal as it is thought that photoevaporation inflates discs around low-mass protostars \citep{jcrev}. These effects would only be enhanced for discs subject to more ionising photons which is more likely from MYSOs. Flaring could be indicative of relatively higher disc temperatures caused by lower opacity (lower disc densities) and increased irradiation of the disc. This implies that the disc density is lower and that more energetic stellar irradiation is present and that photoevaporation could be occurring \citep{jcrev}. 

The presence of multiplicity cannot be ruled out, however. Recent work has shown that a large proportion of MYSOs are expected to exist in multiple systems \citep{sausagebinary}. IRAS 17216-3801 is the only confirmed binary in the present sample \citep{kraus17}, and is classified as a Type III source. Further spectrointerferometric measurements at $H-$ or $K$-band wavelengths which could directly probe the inner rims of the discs and the photospheres of the protostars in this sample could allow the direct detection of binary systems, as has been used in works such as \citet{orion}. Work by \citet{oliva} has shown that spiral structures (such as those included in the final model of NGC 2264 IRS1) can appear due to disk fragmentation/binary formation. These are most prominent after timescales up to 15,000 years and past 15,000 years the spiral density perturbations become less pronounced, however we note that their simulations do not include outflows nor consider that the disc could be being continually fed by the surrounding envelope which could sustain such structure. Additionally, we note that theoretical work has shown through hydrodynamical modelling that the inner disc is removed for low-mass/Herbig binary systems over timescales of $\sim$0.5Myrs \citep{price18} but such clearing is not apparent in current fragmentation models such as those done by \citet{oliva}. Follow-up observations, such as image reconstruction at 0.01" scales using the VLTI, would allow these cleared inner regions to be investigated in more detail. In particular, observations with GRAVITY or MATISSE which operate at shorter wavelengths and could directly probe these inner regions. %Modelling of massive systems specifically would reveal whether this timescale decreases, as the overall formation timescale does, for massive protostars.

The processes associated with asymmetric substructures in low-mass discs, like the spiral/gap structure inferred around NGC 2264 IRS1, are the presence of binary/multiple systems, dust trapping/growth \citep{zhang15} and the presence of planets \citep{pinte}. Given %that increasing the fraction of large dust grains in the disc improved the fit of the MIDI data for this source, and 
the high multiplicity fraction for MYSOs at wide separations  \citep{sausagebinary}, we conclude that dust growth and the presence of a binary/multiple system are the most likely candidates for any substructure in this disc. Both the effects of binaries and dust growth take tens of thousands of years to occur, implying that NGC 2264 IRS1 may be a more evolved source which is in agreements with its classification as a Type III source. 

\section{Conclusions} \label{sec:conclusions}

Through the unique combination of a multi-scale study of the physical nature of MYSOs and near-infrared spectroscopic work, we have attributed the physical characteristics of a sample of MYSOs to a potential evolutionary sequence based on IR-line emission. The sample constitutes the largest modelled with infrared-derived constraints on physical properties spanning several orders of magnitude in physical size through the simultaneous fitting of interferometric, imaging and SED data. The MYSOs themselves cover a range in suspected masses and luminosities of 8-38 solar masses and $\sim$4000-151000L$_{\odot}$. The disappearance of lines associated with outflow shocks colliding with dense envelope material in the near-infrared spectra and emergence of lines associated with the production of UV photons indicates evolution. In tandem, the sources which display lines the suggest the presence of UV photons are fit with models with low densities through the radiative transfer modelling. This suggests that the envelopes and outflows of the environments are dissipating. Additionally we find that the models of sources who are spectrally classified as more evolved are the only ones to show evidence of substructure. This implies that envelopes and discs are co-evolving for MYSOs, similar to low-mass stars and despite the rapid timescale of massive star formation. Future studies continuing to exploit milli-arcsecond resolution observations on an expanded sample, particularly on younger MYSOs who are in the minority in our sample, will be important aides in confirming these and ultimately probing the physics and chemistry of the disc evolution MYSOs.
%As the MYSOs were all studied using the same methodology, their results could be compared. %Correlations were used to look for trends in the protostellar characteristics and no bias was found towards inclination or the resolving power of the observations. Of the strong correlations, perhaps the most notable is that between minimum dust radius and luminosity. Given that the minimum dust radius is larger than the sublimation radius for half of the sources, this implies that the cause of the clearing is still linked to the luminosity and that perhaps photoevaporation is the culprit. Comparison of our work to that of \citet{coop} allowed the physical parameters of the MYSOs to be put into an evolutionary perspective. 
%We find that all the Type III sources (the most evolved sources), display lower densities, larger cavity opening angles, a large degree of disc flaring and all display some level of substructure in their disc. This is consistent with the idea that substructure evolves over time, and may have similar origins as observed for low-mass sources, such as disc dispersal.
Amongst the sample, the sources modelled with the largest outflow cavities also fall in the category associated with the latest evolutionary phase. We compare the geometrical characteristics of the sample \citep{frost2} with the results of 3D radiation-hydrodynamical simulations of YSOs to put quantitative values to the ages of the MYSOs based on their cavity opening angles and find an age range across the types from $\sim$10$^{3-5}$yrs.
%, which agree with our classification and agree that the cavity opening angles get wider as MYSOs evolve. 
The presence of substructures within the elder massive young stellar object discs with suspected ages of $\sim$10$^4$-10$^5$yrs implies that disc evolution can occur around massive protostars, despite these fast formation timescales. Follow-up work to investigate the nature of the observed inner holes and structures of these discs can shed light on the mechanisms that form these structures and further improve our understanding on the role of discs in massive star formation.% and disc evolution as a whole. 
%These conclusions could only be drawn because all the objects were modelled and analysed in the same way.  Determining the cause of the holes observed within the evolved MYSO discs presents an avenue of further work. 

\begin{acknowledgements}

This work was published with funds from the European Research Council under European Union’s Horizon 2020 research programme (grant agreement No 772225). We thank the STFC for funding this PhD project and in addition thank the referees and Prof. Melvin Hoare and Prof. Steven Longmore for their suggestions and discussions which enabled the improvement of this work. 

\end{acknowledgements}

\bibliography{sample631}{}
\bibliographystyle{aasjournal}



\end{document}